\documentclass[journal]{IEEEtran}

\ifCLASSINFOpdf
\else
   \usepackage[dvips]{graphicx}
\fi
\usepackage{url}

\usepackage{graphicx}
\usepackage{multirow}

\usepackage{amssymb}
\usepackage{xcolor}
\usepackage{cite}
\usepackage{amsthm}
\usepackage{amsmath}
\usepackage{booktabs}
\usepackage{multirow}
\usepackage{algorithmic}
\usepackage{tabularx}
\usepackage{bbding}
\usepackage{hyperref}
\usepackage{verbatim}
\usepackage[linesnumbered,lined,boxed,commentsnumbered,ruled]{algorithm2e}
\usepackage{subfig}
\usepackage{soul}
\usepackage{amssymb}
\usepackage{pifont}
\newcommand{\cmark}{\ding{51}}%
\newcommand{\xmark}{\ding{55}}%

\begin{document}

\title{Context-Aware Two-Step Training Scheme for Domain Invariant Speech Separation}

\author{Wupeng Wang, Zexu Pan, Jingru Lin, Shuai Wang, \IEEEmembership{Member, IEEE},
Haizhou Li, \IEEEmembership{Fellow, IEEE}} 

\maketitle

\begin{abstract}
Speech separation seeks to isolate individual speech signals from a multi-talk speech mixture. Despite much progress, a system well-trained on synthetic data often experiences performance degradation on out-of-domain data, such as real-world speech mixtures. To address this, we introduce a novel context-aware, two-stage training scheme for speech separation models. In this training scheme, the conventional end-to-end architecture is replaced with a framework that contains a context extractor and a segregator. The two modules are trained step by step to simulate the speech separation process of an auditory system. We evaluate the proposed training scheme through cross-domain experiments on both synthetic and real-world speech mixtures, and demonstrate that our new scheme effectively boosts separation quality across different domains without adaptation, as measured by signal quality metrics and word error rate (WER). Additionally, an ablation study on the real test set highlights that the context information, including phoneme and word representations from pretrained SSL models, serves as effective domain invariant training targets for separation models.
\end{abstract}


\IEEEpeerreviewmaketitle

\section{Introduction}

\IEEEPARstart{S}peech separation is a process of isolating individual speech signals from a multi-talker acoustic environment, which is commonly known as the `cocktail party' problem. It is an enabling technology essential for telecommunication, audio systems, and hearing aids. Moreover, it's crucial for the robustness of downstream tasks, such as automatic speech recognition (ASR)~\cite{wang2018voicefilter,gabrys2022voice, pan2022hybrid}, voice conversion~\cite{zhou2020multi,zhou2021seen}, and speaker recognition~\cite{rao2019target,xu2021target}.

Speech separation techniques, from unsupervised independent component analysis (ICA)~\cite{lee1998independent,araki2004underdetermined,choi2005blind} and non-negative matrix factorization (NMF)~\cite{wang2014discriminative,weninger2014discriminative,virtanen2007monaural}, to supervised end-to-end approaches~\cite{luo2019conv,luo2020dual, subakan2021attention, wang2021neural, zhao2023mossformer, wang2023tf, li2022efficient, pan2021reentry, pan2023imaginenet}, have advanced significantly in the recent past. However, It was observed that the separation quality of these models tends to degrade considerably in face of out-of-domain data, especially real-world speech mixtures. This degradation is likely due to the substantial mismatch between synthetic and real-world mixtures, including variations in accent, intonation, and environmental factors, which is commonly referred to as the `domain-mismatch' problem~\cite{wang2024speech}.  


Being immune from such domain-mismatch, human auditory system could easily capture the target voice even in unfamiliar environments. One possible reason is that human auditory system is capable of following perceptual objects, such as phonemes, words and contextual information, which is inherently domain-invariant. This hypothesis is supported by the computational auditory scene analysis (CASA) theory. In~\cite{bregman1994auditory}, Bregman et al. observe that the auditory system often restores the phoneme information to complete perception, even when it is contaminated by other sudden changes. A similar observation regarding word information is reported in the research of Warren et al.~\cite{warren1970perceptual}. Inspired by these findings, we believe that incorporating perceptual objects can help separation models address the domain mismatch problem, thereby maintaining speech separation quality in out-of-domain scenarios.

\begin{figure*}[t]
	\centering 	\includegraphics[width=0.82\linewidth]{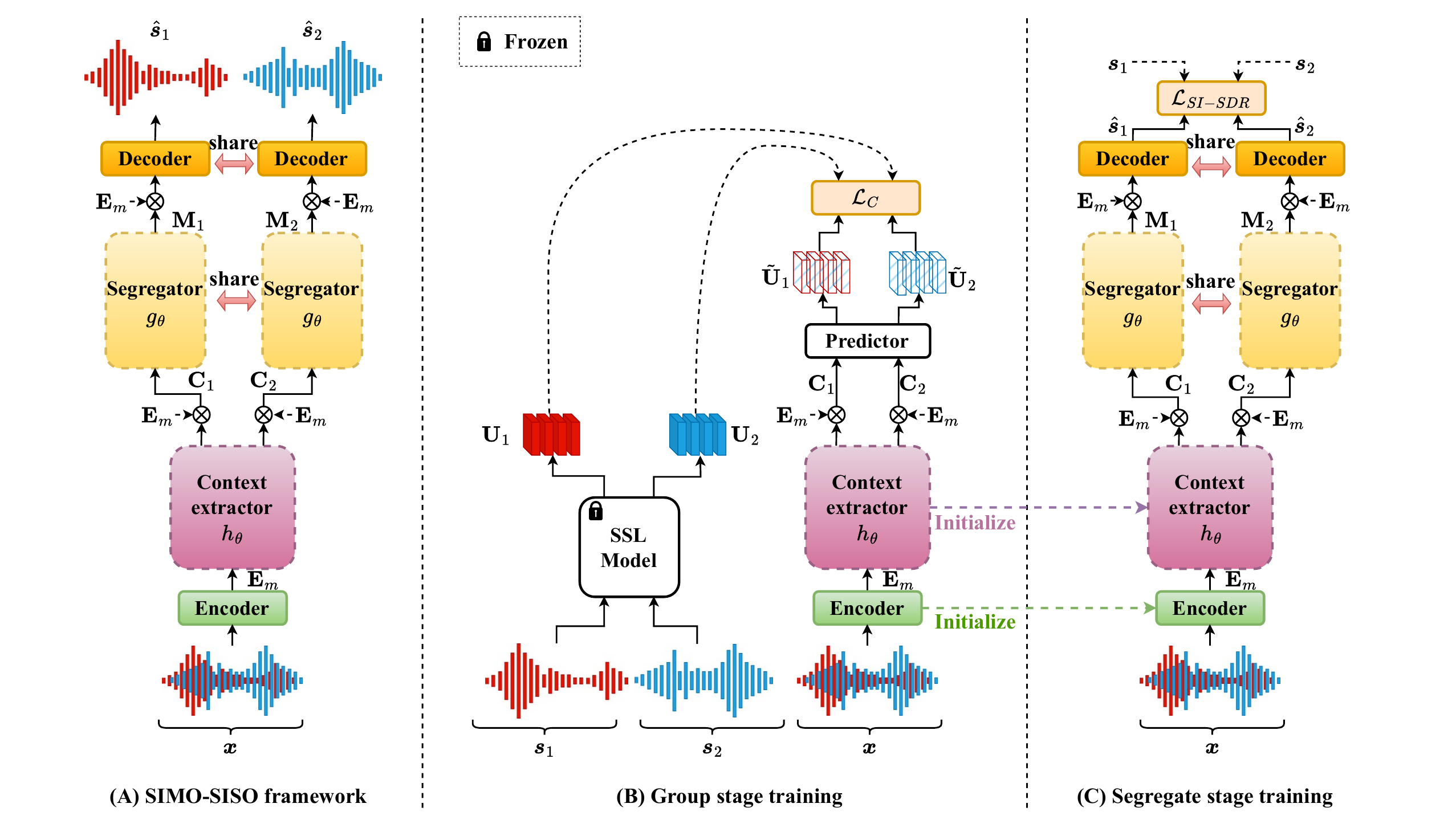}
	\caption{(A) illustrates a general SIMO-SISO framework for any speech separation model. (B) and (C) are schematic diagrams of the two training stages.  $\boldsymbol{x}$ is the mixture waveform.  $\mathbf{E}_m$ is the latent embedding extracted by the encoder. $\mathbf{C}_1$ and $\mathbf{C}_2$ are the predicted contextual embedding.  $\tilde{\mathbf{U}}_1$ and $\tilde{\mathbf{U}}_2$ are the estimated contextual representations, and  $\mathbf{\mathbf{U}}_1$ and $\mathbf{\mathbf{U}}_2$ are the ground-truth contextual representations.  $\mathbf{M}_1$ and $\mathbf{M}_2$ are the latent mask from the segregator $h_{\theta}$. The $\hat{\boldsymbol{s}}_1$ and $\hat{\boldsymbol{s}}_2$ are the estimated signal, and the $\boldsymbol{s}_1$ and $\boldsymbol{s}_2$ are the target reference speech.} 
	\label{Fig.method} 
\end{figure*}

The utilization of perceptual objects has been widely explored for monaural speech processing. In~\cite{wang2016phoneme}, Wang et al. proposed  phoneme-specific deep neural networks to improve speech intelligibility. Schulze et al.~\cite{schulze2020joint} introduced a text-informed phoneme alignment method to capture the temporal relationship between speech signals and their corresponding phoneme labels. Du et al.~\cite{du2020pan} integrated the phonetic posterior-gram information into the enhancement network to iteratively increase the denoising quality. Extending from phonemes to words, Yang et al.~\cite{yang2023selective} implemented a feature-wise linear modulation (FiLM)~\cite{perez2018film} to adaptively perceive word-level information. Despite the convincing advantages of incorporating perceptual objects, these approaches often rely on speech-text parallel data, thereby restricting their applicability in speech separation tasks. 

The self-supervised learning (SSL) pretrained model offers a promising alternative to addressing the out-of-domain issues. The self-supervised models can serve as feature extractors that learn from a large unlabeled dataset to capture the high-level domain invariant contextual information. In~\cite{shi2021discretization}, Shi et al. explored the discretization of separated speech using SSL models, followed by resynthesis with a pretrained vocoder. Tal et al.~\cite{hsieh21improving, tal2022systematic, close2023perceive, lin2024selective, lin24sa} examined the usage of SSL representations as supervision, regularization, and conditioning signals for downstream models.  Wang et al.~\cite{wang2024speech} investigated the integration of a domain-invariant pretrained frontend to improve the separation quality of downstream models in real scenarios. Although accurate text annotations are not required for model training, the need for auxiliary SSL networks still introduces additional computational costs and increases the processing burden. 

In~\cite{bregman1994auditory}, it is suggested that the speech separation procedure of the auditory system generally involves two stages: the group stage that extracts perceptual objects of the input signal, and the segregate stage where these objects are segregated into distinct audio streams. Inspired by this formulation, we propose a context-aware two-stage training scheme to perceive the contextual information without an auxiliary network, thus improving the domain transfer capacity of separation models. The contributions of our paper are summarized as follows,

\begin{itemize}
    \item We introduce a novel two-stage training scheme designed for the perception of contextual information;
    \item We propose a contextual loss to maximize the mutual information between the predicted contextual representations and the  representations derived from SSL models;
    \item We conduct comprehensive experiments to evaluate the effectiveness of contextual information across diverse real and synthetic datasets, with various downstream models and evaluation metrics; 
\end{itemize}


\section{Methdology}
\label{sec:model}
To emulate the perceptual processes of the auditory system, we follow~\cite{luo2021rethinking} to decompose any separation model with $N$ separation blocks into two parts: a single-input-multi-output (SIMO) context extractor $h_{\theta}$ with $\frac{N}{2}$ separation blocks, and a single-input-single-output (SISO) segregator $g_{\theta}$ with another $\frac{N}{2}$ separation blocks. The former mimics the group stage of the auditory system to capture domain-invariant contextual information, while the latter reconstructs the target reference speech for each speaker in an end-to-end manner, which is similar to the segregate stage of the auditory system. Each branch of the segregator shares parameters to prevent auxiliary complexity. 

\newcolumntype{Y}{>{\centering\arraybackslash}X}
\newcolumntype{Z}[1]{>{\centering\arraybackslash}p{#1}}
\begin{table*}
    \centering
    \caption{A comparison among the ConvTasNet with our two-stage training scheme, and its variants. The test sets are from both the matched test set LM2Mix and the mismatched test set LRS2Mix.  ``\cmark''  in the table denotes that the method or framework is implemented, while  ``\xmark''  denotes otherwise.  ``N/A'' means no SSL model is required.  ``TS'' means our two-stage training scheme.  ``Mel'' is the Mel-spectrogram. The \textbf{bold values} are the optimal choices.}
    \begin{tabularx}{\textwidth}{@{}Z{0.8cm}|Z{1.2cm}|Z{0.4cm}|Z{1.4cm}|Z{1.6cm}|Y|Y|Y|Y|Y@{}}
		\toprule
		System & SIMO & \multirow{2}*{TS} & SSL & Group stage & Dev & \multicolumn{2}{c|}{
Test (LM2Mix)} & \multicolumn{2}{c}{Test (LM2Mix $\rightarrow$ LRS2Mix)}\\
        (Sys.) & -SISO & ~ & model & target & SI-SDRi (dB)$\uparrow$ & SI-SDRi (dB)$\uparrow$ & SDRi (dB)$\uparrow$ & SI-SDRi (dB)$\uparrow$ & SDRi (dB)$\uparrow$\\
		\midrule
		Baseline & \xmark & N/A & N/A & N/A & 14.5 & 14.1 & 14.5 & 8.6 & 9.1\\ 
  
        \midrule
        1 & \cmark & \xmark & N/A & N/A & 12.5 & 13.2 & 13.6 & 7.0 & 7.4\\
        
        \midrule
        2 & \cmark & \cmark & N/A & Signal & 15.5 & 14.9 & 15.3 & 8.6 & 9.2\\
        \midrule
        
        3 & \multirow{4}*{\cmark} & \multirow{4}*{\cmark} & \multirow{4}*{Hubert} & Mel & 15.3 & 14.7 & 15.1 & 8.8 & 9.3\\
        4 & ~ & ~ & ~ & Phoneme & 15.0 & 14.8 & 15.2 & 9.0 & 9.5\\
        5 & ~ & ~ & ~ & Word & 15.1 & 14.7 & 15.1 & 9.2 & 9.7\\
        \textbf{6}  & ~ & ~ & ~ & \textbf{Hybrid} & \textbf{15.6} & \textbf{15.0} & \textbf{15.4} & \textbf{9.5} & \textbf{9.9}\\
        \midrule
        7 & \multirow{2}*{\cmark} & \multirow{2}*{\cmark} & Wav2vec2.0 & \multirow{2}*{Hybrid} & 15.5 & 15.0 & 15.4 & 9.2 & 9.7\\
        8 & ~ & ~ & WavLM & ~ & 15.4 & 14.9 & 15.3 & 9.2 & 9.6\\
		\bottomrule
    \end{tabularx}
    \label{tab_ssl}
\end{table*}

\subsection{SIMO-SISO Framework}

{As illustrated in Fig.~\ref{Fig.method} (A), given a time-domain multi-speaker mixture waveform $\boldsymbol{x} \in \mathbb{R}^{\tau \times 1}$ of $\tau$ samples containing two isolated target reference speech $\boldsymbol{s}_1$ and $\boldsymbol{s}_2$, we first encode it into $T$ frames of $D$-dimensional spectrum-like latent embeddings in a sequence  $\mathbf{E}_m \in \mathbb{R}^{T \times D}$ with 1D convolutional encoder followed by a rectified linear activation (ReLU). Then the context extractor $h_{\theta}$ is applied to estimate the contextual embedding $\mathbf{C}_i$ with $i \in \{1,2\} $ representing one of the two speakers. Next, the $\mathbf{C}_i$ serves as the input of the segregator $g_{\theta}$ to predict a latent mask $\mathbf{M}_i$. Finally, we multiply the mixture embedding $\mathbf{E}_m$ with the latent mask $\mathbf{M}_i$ and send it to a 1D convolutional decoder to reconstruct the target reference speech.} 

\subsection{Context-Aware Two-Stage Training Scheme}
We propose a novel two-stage training scheme with a contextual loss $\mathcal{L}_{\text{C}}$ for the SIMO-SISO framework, namely group stage training followed by segregate stage training. 


\subsubsection{\textbf{Group stage}} The first stage in Fig.~\ref{Fig.method} (B) is to capture the contextual information through context extractor $h_{\theta}$. As reported in~\cite{hsu2021hubert}, each layer of a SSL model encodes unique acoustic and linguistic information. 
{For example, the $7^{th}$ layer of the 7-layer convolutional neural network encoder captures information closely aligned with the Mel-spectrogram~\cite{hsu2021hubert, pasad2021layer, pasad2023comparative}. In Hubert and Wavlm, the $11^{th}$ and the $9^{th}$ layers of the 12-layer transformer context network mainly correspond to phoneme and word-related features, respectively, as indicated by the probing analysis in~\cite{pasad2021layer}, while for Wav2vec2.0, the $6^{th}$, and the $8^{th}$ transformer layers are notably associated with phoneme and word-level information~\cite{pasad2023comparative}.

Motivated by this, we extract three representations: Mel-spectrogram, phoneme, and word-related representations, from specific layers of SSL models to serve as the training target $\mathbf{U}_i$,  $i \in \{1,2\}$. The phoneme and word-related representations are seens as context aware training targets. We train the context extractor $h_{\theta}$ to generate $\tilde{\mathbf{U}}_i$ by a shared 1D convolutional predictor. The training process employs permutation invariant training (PIT)~\cite{yu2017permutation} with a contextual loss $\mathcal{L}_{\text{C}}$ as follows:}
\begin{equation}   
\label{eqa.infonce}
\mathcal{L}_{\text{C}}\left(\mathbf{U}_i, \mathbf{\tilde{U}}_i\right) = -\log \frac{\exp\left(\psi\left(\mathbf{U}_i, \mathbf{\tilde{U}}_i\right) / \omega\right)}{\sum_{\mathbf{U} \in \mathcal{Q}_\mathbf{U}} \exp\left(\psi\left(\mathbf{U}, \mathbf{\tilde{U}}_i\right) / \omega\right)},
\end{equation}
{The $\mathcal{L}_{\text{C}}$ is the InfoNCE~\cite{oord2018representation} loss to maximize the mutual information between predicted contextual representations $\tilde{\mathbf{U}}_i$ and training target $\mathbf{U}_i$ through contrastive learning. $\mathbf{U}_i$ here is one of the ground-truth contextual representations from SSL models.} The $\omega$ is the temperature and the $\psi$ stands for the similarity function to distinguish the ground-truth contextual representations $\mathbf{U}_i$ from a set of candidates $\mathcal{Q}_\mathbf{U}$, which includes $\mathbf{U}_i$ and other negative distractors. 

\subsubsection{\textbf{Segregate stage}}The second stage in Fig.~\ref{Fig.method} (C) generates the isolated audio streams through the entire separation model. We initialize the context extractor with weights obtained from the group stage and jointly train the model with signal similarity loss $\mathcal{L}_{\text{SI-SDR}}$ introduced in~\cite{luo2019conv}, to reconstruct the target reference speech. The SSL model and predictor used during the group stage are discarded in this stage. 


\section{Experiment}
\label{sec:exp}
\subsection{Experiment Setup}

To investigate the domain transfer capacity of our proposed two-stage training scheme, we follow~\cite{wang2024speech} to train the separators with the Libri2Mix dataset~\cite{cosentino2020librimix} (LM2Mix), test them on synthetic test sets from LM2Mix and LRS2Mix dataset~\cite{li2022efficient}, and the real test set from Real-M dataset~\cite{subakan2022real}. The LM2Mix dataset is simulated at a 16kHz sampling rate based on Librispeech (LS) corpus~\cite{panayotov2015librispeech}. The train set contains 13,900 utterances from 251 speakers, while the development set includes 3,000 utterances from 40 speakers. All utterances are randomly selected from the LibriSpeech corpus and mixed by specific loudness units relative to full scale. As for the test set, there are 3,000 utterances simulated by the speech voices of 40 speakers. The test set of LRS2Mix~\cite{li2022efficient} contains 3,000 utterances generated from the spoken sentences audio recordings from BBC television~\cite{chung2018voxceleb2} with signal-to-noise ratios sampled between -5 dB and 5 dB. For the real test set, we utilize 837 samples from the REAL-M, which is recorded by asking contributors to read a predefined set of sentences simultaneously in complex acoustic environments.

 	

The implementation of ConvTasNet~\cite{luo2019conv} and DPRNN~\cite{luo2020dual} are based on the Asteroid toolkit\footnote{\url{https://github.com/asteroid-team/asteroid}}, while for Mossformer~\cite{zhao2023mossformer}, we integrate the official implementation from the SpeechBrain\footnote{\url{https://github.com/speechbrain/speechbrain} toolkit into the Asteroid toolkit to ensure a fair cross-model comparison. The stride and the kernel size of the encoder and decoder are 16 and 32, respectively. Other parameters are similar to the official configuration in the asteroid and speechbrain toolkits.} We train all models with official recipes. The batch size of all separation models is 2 and the maximum training epochs is 200. The learning rate is initialized as 1e-3 and decays half when the loss on the validation set is not improved in 5 consecutive epochs. Early stopping is applied if no better model is found in the validation set for 30 consecutive epochs. We utilize the Adam optimizer for back-propagation of the separation model during the training stage. The similarity function $\psi$ is the cosine similarity, and the temperature is set to $0.1$. 

\subsection{Experiment Result}

We first investigate the effectiveness of our proposed two-stage training scheme on synthetic datasets. The results are illustrated in Table.~\ref{tab_ssl}. The baseline system is the ConvTasNet trained in an end-to-end manner, which achieves a performance of 14.1 dB SI-SDRi and 14.5 dB SDRi on the LM2Mix test set, demonstrating strong performance in scenarios where there is no domain mismatch between training and testing. However, directly applying the model to the mismatched LRS2Mix test set (LM2Mix $\rightarrow$ LRS2Mix), the performance drops to 8.6 dB in SI-SDRi and 9.1 dB in SDRi. These results indicate that despite both LM2Mix and LRS2Mix being synthetic datasets, the domain gap substantially degrades separation quality. 



In Sys.1, we implement the SIMO-SISO framework and train it using a conventional end-to-end approach. Due to the absence of group stage targets, the separation quality significantly degrades, indicating that simply replacing the baseline with the SIMO-SISO framework does not boost performance. We then adopt the SI-SDR loss as the training target during the group stage. As shown in Sys.2, this approach achieves 14.9 dB in SI-SDRi and 15.3 dB in SDRi on the in-domain LM2Mix test set. However, when applied to the out-of-domain LRS2Mix test set, the separation quality remains comparable to that of the baseline system. This finding suggests that while signal-level, domain-specific knowledge improves in-domain separation, it is insufficient for addressing domain mismatch in separation models. Next, we incorporate the Mel-spectrogram, phoneme, and word representations extracted from Hubert as the training target in the group stage. As demonstrated in Sys.3, leveraging low-level Mel-spectrogram features offers marginal improvement in separation performance in out-of-domain scenarios. In contrast, using high-level phoneme (Sys.4) and word (Sys.5) representations boosts separation quality on the LRS2Mix test set. The best choice is using the word representation as the training target in the group stage, achieving 9.2 dB in SI-SDRi and 9.7 dB in SDRi on the LRS2Mix test set. These results support our hypothesis that domain-invariant contextual information is critical for improving separation quality in out-of-domain scenarios.

Building on these findings, we employ a hybrid loss that combines both signal and all contextual losses, to capture domain-specific and invariant knowledge simultaneously. As shown in Sys.6, the separation quality comes to 15.0 dB in SI-SDRi and 15.4 dB in SDRi on the LM2Mix test set, while for the LRS2Mix test set, it is 9.5 dB SI-SDRi and 9.9 dB SDRi, respectively. These results reveal that the perception of domain-specific and domain-invariant knowledge is beneficial for each other. In Sys.7 to Sys.8, we implement our method with Wav2vec2.0~\cite{baevski2020wav2vec}, and WavLM~\cite{chen2022wavlm}. The results indicate that our proposed method is highly compatible with various SSL models to boost the domain adaptation capability of separation models. Since the Hubert-based hybrid method shows the best results, we select HuBERT as the default group-stage setup in subsequent experiments.


\begin{table}[]
    \caption{A comparison among different separators to transfer the separation knowledge from the LM2Mix dataset to the LRS2Mix test set. All separators are trained in the Sys.6 setting. ``TS'' means our two-stage training scheme. The \textbf{bold values} are the optimal choices.}
    \centering
    \begin{tabular}{c|c|c|c|c} 
       \toprule
        \multirow{2}*{Separator} & SIMO & \multirow{2}*{TS} & \multicolumn{2}{c}{Test (LM2Mix $\rightarrow$ LRS2Mix)}\\
        ~ & -SISO & ~ & SI-SDRi(dB) & SDRi(dB)\\
        
        \midrule    
        \multirow{2}*{ConvTasNet~\cite{luo2019conv}} & \xmark & N/A & 8.6 & 9.1 \\
        ~ & \textbf{\cmark} & \textbf{\cmark} & \textbf{9.5} & \textbf{9.9} \\

        \midrule    
        \multirow{2}*{DPRNN~\cite{luo2020dual}} & \xmark & N/A & 9.3 & 9.8\\
        ~ & \textbf{\cmark} & \textbf{\cmark} & \textbf{9.7} & \textbf{10.1} \\

        \midrule    
        \multirow{2}*{Mossformer~\cite{zhao2023mossformer}} & \xmark & N/A & 12.2 & 12.6  \\
        ~ & \textbf{\cmark} & \textbf{\cmark} & \textbf{12.7} & \textbf{13.1}\\
       \bottomrule
    \end{tabular}
    \label{tab_sep}
\end{table}

The proposed framework is applicable to any speech separation models. Next, we validated its generalizability across different separator backbones by training them on the LM2Mix dataset and testing them on LRS2Mix without any adaptation. As shown in Table.~\ref{tab_sep},  all three separators with our proposed two-stage training scheme outperform their original settings that are trained in an end-to-end manner, despite that the separators have not seen the labeled data from the LRS2Mix dataset. These results suggest that the proposed framework and two-stage training scheme can be generalized to different separation models for out-of-domain scenarios.

\begin{figure}[]
    \centering 
    \includegraphics[width=0.8\linewidth]{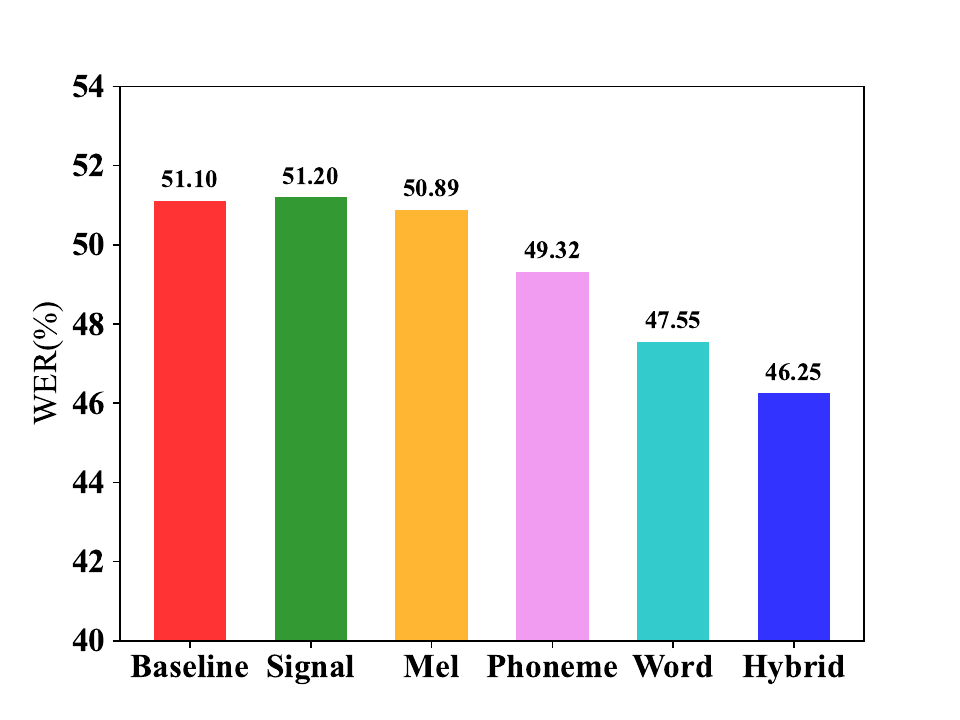}
    \caption{The automatic speech recognition results in terms of word error rate (WER \%) for  ConvTasNet separator that is trained on the LM2Mix with various supervisory targets and tested on the REAL-M test set.} 
    \label{Fig.convtasnet} 
\end{figure}

Finally, we explore the effectiveness of our method in real-world scenarios. We conduct an ablation study using ConvTasNet, a model widely adopted in both academic research and industrial applications due to its low computational requirements. 

As illustrated in Fig.\ref{Fig.convtasnet}, the ConvTasNet trained with our approach boosts separation quality in terms of word error rate (WER) when evaluated by a pretrained transformer-based speech recognition model from ESPNet~\cite{watanabe2018espnet}. The observed trend aligns with the results of the LRS2Mix synthetic test set. For instance, the end-to-end ConvTasNet achieves a 51.08\% WER on the REAL-M dataset. Using the speech signal as the target during the group stage does not boost recognition accuracy. Employing the Mel-spectrogram, phoneme, and word representations as training targets 
improves separation quality, while the hybrid target achieves the lowest WER (46.25\%). These findings further validate that contextual information in group stage training is effective in enabling speech separation models to generate high-quality separated speech in real-world scenarios.

\section{Conclusion}
\label{sec:con}
{We introduced a novel, context-aware, two-stage training scheme for speech separation. We found that we could leverage contextual representations from SSL models to mitigate domain mismatch issues, and overcome domain mismatch.  As validated in a number of experiments, we confirmed that optimizing separation networks with contrastive learning-based contextual loss is beneficial for the generalization capability of separation models in real-world scenarios. Furthermore, our work opens a new direction to overcome the domain-mismatch problem through knowledge transfer.}


\bibliographystyle{IEEEtran}
\bibliography{Bibliography}

\end{document}